\documentclass[aps,preprint,showpacs,superscriptaddress]{revtex4}

\usepackage{graphicx}
\usepackage{amsmath}
\usepackage{bm}

\newcommand{\be}{\begin{equation}}
\newcommand{\ee}{\end{equation}}

\begin{document}

\title
{Oscillations of atomic nuclei in crystals}

\author{V.A. Vdovenkov}

\affiliation{Moscow State Institute of Radioengineering,
 Electronics and Automation (technical university)\\
Vernadsky~ave. 78,~117454~Moscow,~Russia}

\begin{abstract}
Oscillations of atomic nuclei in crystals are considered in this
paper. It is shown  that elastic nuclei oscillations relatively
electron envelops (inherent, I-oscillations) and waves of such
oscillations can exist in crystals at adiabatic condition. The
types and energy quantums of I-oscillations for different atoms
are determined. In this connection the adiabatic crystal model is
offered. Each atom in the adiabatic model is submitted as
I-oscillator whose stationary oscillatory terms are shown as deep
energy levels in crystals. The I-oscillations can be created at
the expense of recombination energy of electrons and holes on
electron-vibrational centers in semiconductors. They interact
among themselves, with phonons and electrons, they influence on
physical properties of crystals and crystal structures. The
I-oscillations representing oscillations of atomic nuclei
relatively electron system in crystals or molecules are the
important physical reality.
\end{abstract}



\maketitle

\section{Introduction}

Electronic properties of crystals (and molecules) usually
investigate by using  of approached  solutions of Schrodinger
equation. The important results of such investigations are
properties of electronic system but the available information
about  energy quantum and types of nuclei oscillations are limited
because of features of used adiabatic approximations. An
information about crystal oscillations usually receive from the
dynamic analysis of traditional crystal models in which  atoms are
replaced by physical points with masses of atoms. Such traditional
models contradict adiabatic theories because nucleus and electron
envelop of every atom in  this model  is one particle and
adiabatic nucleus oscillations (relatively electron environment)
are impossible. Information about nuclei oscillations which was
received on basis of traditional crystal models  is incomplete. At
the same time the role of nuclei oscillations   in various
physical effects, such as phonon drag of electrons and
hyperconductivity - superconductivity at room and at higher
temperatures, is important. Scientific understanding of such
physical effects is inconvenient without enough information about
nuclei oscillations and interaction of these oscillations with
electrons and phonons. In this connection it is useful to consider
the nuclei oscillations in crystals in accordance with adiabatic
theory of solids, to construct the adiabatic crystal model in
agreement with adiabatic theories, to estimate possible influence
of nuclei oscillations on physical properties of crystals or
molecules. Given  paper is devoted to these questions.

\section{Theoretical preconditions}

The modern theory of solids contain the representation about
atomic structure of substances and Schrodinger equation:
\begin{equation}\label{E1}
  i\hbar \frac{\partial\Psi}{\partial t} = H \Psi.
\end{equation}
Hamiltonian (H) for a crystal  include the kinetic energy
operators for nuclei ($T_{z}$) and for electrons  ($T_{e}$), and
also crystal potential (V):
\begin{equation}\label{E2}
  H = T_{z} + T_{e} + V.
\end{equation}

Crystal potential V usually include the operators of Coulomb
interaction between electrons $(V_{e})$, between nuclei $(V_{z})$,
and between electrons and nuclei $(V_{ez}):V = V_{e} + V_{z} +
V_{ez}$. Solution of Eq. (\ref{E1}) is the wave function of a
crystal $\Psi(r,R,t)$ where t - time, r and R designate degrees of
freedom for electronic system  and  for nuclei system,
accordingly. It is impossible to solve precisely the given
equation dependent on huge amount of (more than $10^{23}$)
variables because of known fundamental and technical difficulties.
Theory searches for the reasonable approached solutions of the Eq.
(\ref{E1}) which would allow to calculate and to predict noticed
on experience physical magnitudes, and crystals properties. In
this connection apply an approached methods. So, division of
variables in the Eq.~(\ref{E1}) would allow to simplify the
problem and to consider motion of electron system irrespective of
nuclei system. In such case these systems would submit to
micro-canonical distribution, and the exchange of energy between
systems would be impossible, as though they were divided by
adiabatic membrane. For the first time P. Dirac  has applied
separation of electronic and nuclear variables \cite{Dir30} and
has presented wave function as $\Psi(r,R,t) =
\phi(r,t)\cdot\chi(R,t)$. He show that, supposing the energy
conservation, it is possible to replace the Eq.~(\ref{E1}) by two
equations for wave functions $\phi(r,t)$ and $\chi(R,t)$, which
describe movements of electronic and nuclear systems of a crystal:
\begin{equation}\label{E3}
  i \hbar\frac{\partial\phi}{\partial t} = -
\sum_{i}\frac{\hbar^{2}}{2m}\nabla_{i}^{2}\phi + \{\int dR \cdot
\chi^{*}(R_{I}, t) \cdot V(r_{i},R_{I}) \cdot \chi(R_{I},
t)\}\phi,
\end{equation}
\begin{equation}\label{E4}
  i \hbar\frac{\partial\chi}{\partial t}=-\sum_{I}\frac{\hbar^{2}}{2M_{I}}\nabla_{I}^{2}\chi +
 \{\int dr \cdot \phi(r_{i}, t) \cdot H_{e}(r_{i}, R_{I}) \cdot \phi(r_{i},t)\}\chi,~~
\end{equation}
where m - electron mass, $M_{I}$ - mass of I-th nuclei, $V(r_{i},
R_{I})$ - crystal potential and $H_{e}(r_{i}, R_{I})$ -
Hamiltonian of electron system. Coupled equations
(\ref{E3})-(\ref{E4}) introduce the basis of the time-dependent
self-consistent field (TDSCF) method. One can see from the
Eqs.~(\ref{E3})-(\ref{E4}) that the potential field, in which each
system of particles move, depends on time and from results of
averaging on coordinates of other system. In it the new quality
consists which electrons get at the presence of nuclei. That is
visible even on example of hydrogen atom, where proton and
electron separately have coulomb fields, but in envelop of
hydrogen atom the electron move in coulomb field of nucleus, and
the nucleus move in a field of electronic shell, which have not
coulomb field. The electronic envelops of atoms can unite among
themselves, forming molecules and crystals but electrons outside
of shells can not do that. So the electronic envelops represent
new collective quality of electrons in atom but exact division of
variables  in the Eq.~(\ref{E1}) cannot be carried out, and the
problem about movement of nuclei system  and electron system  in a
crystal generally is self-consistent  and nonadiabatic.
Nevertheless adiabatic approximations are used. Born-Oppenheimer
adiabatic approximation \cite{Born27} is related to the solution
of the stationary Schrodinger equation
\begin{equation}\label{E5}
   H\Psi = W\Psi,
\end{equation}
where W - energy of a crystal. Wave function of a crystal is
product of electronic and nuclear wave functions:
\begin{equation}\label{E6}
  \Psi(r, R) =\varphi(r,R)\cdot\Phi(R).
\end{equation}
One can receive from the equation Eq.~(\ref{E5}) the following two
equations
\begin{equation}\label{E7}
 (T_{e}+V) \varphi(r, R) = E \varphi(r, R),
\end{equation}
\begin{equation}\label{E8}
  (T_{z} + E + A)  \Phi(R) = W \Phi(R),
\end{equation}
where E - energy of electron system and
\begin{equation}\label{E9}
  A=-\sum_{I}(\hbar^{2}/2M_{I}) \int \varphi^{*}(r, R) \nabla^{2}_{R}
  \varphi(r, R)\cdot d \tau_{r}
\end{equation}
- adiabatic potential.  Equations (\ref{E7}, \ref{E8}) form the
connected system because potential A includes dependence of
electronic wave function from change of nuclei coordinates  and
crystal potential V depends from r and R. In it the reasons of
unadiabatic interaction of electrons system  with nuclei system
consist. Potential A describe stationary and, hence, oscillating
process of energy exchange between systems of electrons and
nuclei. However adiabatic conditions are provided with the certain
accuracy, if potential A is small or is equal to zero. The
Eqs.~(\ref{E6}-\ref{E8}) under condition of A = 0 are known as
Born-Oppenheimer adiabatic approximation. The accuracy of the
given approximation is determined by the contribution of potential
A in energy of a crystal. The appropriate contribution of
displacements, rotations and oscillations of nuclei system as a
power series of small parameter $\eta = (m/M)^{1/4}$ was
calculated in Ref. \cite{Born27}. In accordance with \cite{Born27}
the energy of nuclei oscillations is proportional $\eta^{2}$. In
this connection sometimes assert, that the small magnitude of the
relation (m/M) ostensibly is a condition of a validity for using
of adiabatic approximation. This opinion is inexact, as the
relation (m/M) is always small, but the validity of adiabatic
approach depends on physical conditions and not always is
justified. The wave function in Eqs.~(\ref{E6}) generally speaking
differs from the exact solution and after her substitution in the
Eqs.~(\ref{E5}) the additional components turn out due to which
the movements of nuclei are capable to cause electronic
transitions and by that to break adiabatic condition. C. Herring
has shown \cite{Herr56}, that adiabatic approximation is correct,
if
\begin{equation}\label{E10}
  E_{ij} \gg \sum_{\nu}\hbar\omega_{\nu} R_{\nu}\int d^{3}r\varphi_{i}^{*}
 \frac{\partial}{\partial R_{\nu}}\varphi_{j},
\end{equation}
where $E_{ij}$ - energy of enabled  electronic transition between
conditions i and j, $\nu$ - number of oscillation modes for nuclei
system, $dR_{\nu}$ - characteristic displacement of nuclei system
at frequency with number $\nu$. In case of crystals the systematic
displacement and rotations of all system of nuclei can be
excluded. It may to take into account only oscillations of nuclei
at considering fixed crystal for estimation of adiabacity. Then in
agreement with \cite{Dav73} the sufficient condition for validity
of adiabatic approach on the only frequency with number $\nu$ is
smallness of nuclei oscillations energy in comparison with
$E_{ij}$:
\begin{equation}\label{E11}
  E_{ij} \gg \hbar \omega_{\nu}.
\end{equation}
Believing that the conditions Eqs.~(\ref{E10})-(\ref{E11}) are
satisfied, sometimes, one apply additional approximation   in
which wave function of a crystal (for estimation of potential A
and for calculation of energy for electronic system E)  is product
of independent from each other wave functions of electronic and
nuclear systems:
\begin{equation}\label{12}
   \Psi(r, R) = \varphi(r,R_{0}) \cdot \Phi(R),
\end{equation}
where $R_{0}$ - set of nuclei equilibrium positions.

Thus, the stationary oscillations of nuclei system are possible at
adiabatic approximation conditions. Adiabatic oscillations of
nuclei system do not depend on oscillations of electronic system.
They do not influence electronic system and do not change
electronic structure of a crystal as a whole. From the point of
view of adiabatic theories the crystal lattice is formed by system
of electrons, and the nuclei are placed in potential minima of
electronic system, where they can carry out oscillations
concerning electronic system. The opportunity of nuclei
oscillations becomes a reality when the necessary energy is
transmitted to them. These conditions can be executed, for
example, with the help of the electron-vibrational centers (EVC).
Adiabatic oscillations of nuclei are important but for the present
time they are a poorly investigated type  of oscillations in
crystals and molecules.

\section{Oscillations of atomic nuclei}

Solutions of the Eqs.~(\ref{E7}) and (\ref{E8}) in adiabatic
approximation  do not depend from each other. The analysis of
nuclei system adiabatic oscillations may be carried out by
analyzing the solutions of Eq.~(\ref{E8}). This many particle
problem about movements of nuclei system is possible to reduce,
for example by Hartree method \cite{Har28}, to one particle
problem about movement of one j-th  nucleus in effective potential
field $V(R_{j})$ depending from coordinates only of j-th nucleus:
\begin{equation}\label{E13}
  {T_{j}+V(R_{j})}F_{j} = W_{j} F_{j},
\end{equation}
where $T_{j}$ - operator of kinetic energy and $F_{j}$ - wave
function, $W_{j}$ - energy of stationary oscillations of j-th
nucleus in a potential field $V(R_{j})$. This field is created by
all electrons and nuclei of a crystal besides the j-th nucleus.
The electronic environment of j-th atom brought  basic
contribution in $V(R_{j})$ . Contribution of electrons and nuclei
from other atoms of a crystal is insignificant because of their
symmetric and removed position concerning  j-th atom nucleus.
Adiabatic oscillations of nucleus occur on frequencies with number
$\nu$ (which, generally speaking, differ from characteristic
frequencies of electronic system oscillations) near to a minimum
$V(R_{j})$ conterminous with the center of atom electronic shell.
They do not cause an electronic transitions and the electronic
environment of each atom remains constant and motionless on
frequencies with number $\nu$. It is possible to name these
oscillations as inherent or I-oscillations, because of their
properties are determined by inherent parameters of atom: mass and
charge of a nucleus, potential $V(R_{j})$ near to the center of an
electronic envelop.

\begin{figure}
\vspace*{0cm}
\includegraphics[width=12cm]{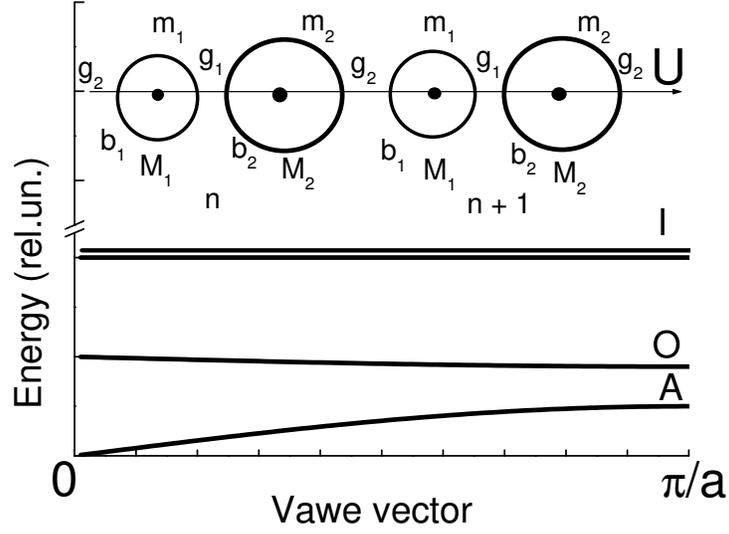}
\vspace*{0cm} \caption {Dispersive  branches of acoustical (A),
optical (O) and inherent (I) oscillations for linear two-nuclear
chain according to Eq. (\ref{E18}). A fragment of such chain
(cells with numbers n and n+1) is shown in the top part of the
figure.} \label{fig1}
\end{figure}

Accordingly each atom in adiabatic models of a crystals should be
presented  as inherent oscillator (I-oscillator) consisting from a
nucleus and electronic shell, which is  connected with each other
by elastic force. In such case unadiabatic processes are
transitions between stationary conditions of adiabatic model. The
electron-vibrational transitions caused by presence of potential A
in the Eq.~(\ref{E8}) are a typical example  of unadiabatic
processes in which I-oscillations of the electron-vibrational
centers, electrons and phonons strongly interact among themselves.
Thus the I-oscillator energy levels (I-terms) of the centers are
shown as deep energy levels. Nuclei oscillations relatively
electrons shells of atoms  in crystal occur in small area $(\simeq
10^{2}\AA)$ and in general they are necessary  to research by
quantum methods. Nevertheless in classical molecular dynamics the
nuclei movements describe  by Newtonian equations  of motion. If
to study harmonic oscillations then it is suitable to use the
known correspondence between results of quantum and classical
theories. This correspondence consist in coinciding the transition
frequency between adjacent quantum oscillatory levels of harmonic
oscillator with his classical oscillation frequency. So energy
spectrum of harmonic I-oscillations is possible to research by
classical method. We used this possibility for description of
oscillations in adiabatic crystal model when nuclei and electron
envelops can move along straight axis U. Such diatomic model is
shown in the top part of Fig.~\ref{fig1} where circles are
electronic envelops and closed points in circles centers are
nuclei of atoms. System equations of motion for this model may be
written:
\begin{equation}\label{E14}
M_{1} \frac{\partial^{2}}{\partial
t^{2}}U_{1n}^{'}=-b_{1}(U_{1n}^{'}-U_{1n}^{''}),~~~~~~~~~~~~~~~~~~~~~~~~~~~~~~~~~~~~~~~~~~~~~
\end{equation}
\begin{equation}\label{E15}
M_{2}\frac{\partial^{2}}{\partial
t^{2}}U_{2n}^{'}=-b_{2}(U_{2n}^{'}-U_{2n}^{''}),~~~~~~~~~~~~~~~~~~~~~~~~~~~~~~~~~~~~~~~~~~~~~
\end{equation}
\begin{equation}\label{E16}
m_{1}\frac{\partial^{2}}{\partial t^{2}}U_{1n}^{''}=-b_{1} (
U_{1n}^{''}-U_{1n}^{'})-g_{1}(U_{1n}^{''} -
U_{2n}^{"})-g_{2}(U_{1n}^{'} - U_{1n-1}^{"}),
\end{equation}
\begin{equation}\label{E17}
m_{2}\frac{\partial^{2}}{\partial
t^{2}}U_{2n}^{''}=-b_{2}(U_{2n}^{''}-U_{2n}^{'})-g_{1}(
U_{2n}^{''}-U_{1n}^{''})-g_{2}(U_{2n}^{''} - U_{1n+1}^{''}),
\end{equation}
where $M_{1}$ and $M_{2}$ - nuclei masses, $m_{1}$ and $m_{2}$ -
an electronic envelops fictitious masses, $U^{'}$ and $U^{''}$ -
displacements of nuclei and electronic shells from equilibrium
position , $b_{1}$, $b_{2}$,  $g_{1}$, $g_{2}$ - coefficients of
elastic forces, t - time, n - elementary cell number. If to be
interested in waves of harmonical oscillations in a linear atomic
chain consisting of atoms which differ only by factors $g_{1}$ and
$g_{2}$, it is necessary to put $m_{1} = m_{1} = m$ , $M_{1} =
M_{2} = M$, $b_{1} = b_{2} = b$. Let's accept for definiteness
$g_{1} > g_{2}$. Then the following dispersion relations are true:
\begin{equation}\label{E18}
  \omega_{1,2,3,4}^{2}(q) = (D/2) \pm (D^{2}/4-F)^{1/2},
\end{equation}
where $D=[M(b+g_{1}+g_{2}-C)-mb](mM)^{-1}$; $F = [b(g_{1} +g_{2} -
B)](Mm)^{-1}$, $B=-[g_{1} + g_{2} + 2g_{1}g_{2}-C]^{1/2}$,
$C=1-2\sin(aq/2)$, q - wave vector, a - lattice spacing. The
dispersive curves qualitatively are represented in Fig.~\ref{fig1}
and contain known acoustic (A) and optical (O) branches, and also
new branches (I) which concern to inherent oscillations
representing oscillations of nuclei relatively electronic
envelops.
In this connection it is expedient to determine magnitudes of
elementary quantums for I-oscillations. Nucleus of j-th atom (or
ion) of crystal oscillate in effective field $V(R_{j})$. It is
obvious that the basic contribution in $V(R_{j})$ introduce the
electrons from envelop of considered j-th atom  but contribution
of nuclei and electronic envelops both environments of the next
and more distant atoms is insignificant because of their symmetric
and distant location relatively of j-th atom. This feature
together with deduction of adiabatic theory allow  to calculate
frequencies of nucleus I-oscillations in a field of his motionless
electronic shell. The potential function $V(R_{j})$ depends mainly
on s-electrons density near  the center of electronic envelop and
in first approximation is spherical symmetric. Therefore small
oscillations of nucleus near to the center of electronic envelop
($\alpha$-type of I-oscillations) practically do not depend on
polarization and have small unharmonicity. Quantums of
I-oscillations in a crystal can be calculated when  the function
$V(R_{j})$ is known. So, elementary quantum of harmonic inherent
oscillations  in statistical model of Thomas-Fermi atom does not
depend on nuclear number and is equal to 0.43~eV. This energy can
be considered as majorization of elementary quantum for
I-oscillations in multi-electronic atom. Function $V(Rj)$ can be
found for any atom from solving the Poisson equation  with
electrical charge density created by electrons of envelop near to
her center. In atom of hydrogen the electronic density is defined
by wave function: $e|\Psi_{1s}|^{2}$, where e - electron charge.
The appropriate magnitude of quantum for harmonic inherent
oscillations $\hbar\omega_{1}$ = 0.519~eV. Un-harmonic correction
to $\hbar\omega_{1}$ in the first and second orders of the
perturbation theory  in condition with oscillatory number $ \nu$~
=~0 is equal to +7~meV (2.7 percent), in condition with $\nu$~=~1
correction is equal to +35~meV (4.5 percent) and in a condition
with $\nu$~=~2 correction is equal  to +70~meV (5.4 percent). The
similar calculations for helium give magnitude of quantum $\hbar
\omega_{2}$ = 0.402~eV. Electronic density near to the center of
electron envelop in multi-electronic atoms is created basically by
s~-~electrons. Quantums of inherent oscillations ($\alpha$ - type
of I-oscillations) for atoms with numbers $Z > 80$ were calculated
taking into account the s-conditions and  the shielding a nuclei
by electrons. Quantums of $\alpha$ - type I-oscillations for atoms
with $Z > 2$ may be determined with help of the following formula
\begin{equation}\label{E19}
\hbar\omega_{z} =
\hbar\omega_{2}\{(\frac{Z-5/16-\xi}{Z-5/16})^{3}\cdot\frac{\Xi(Z-\xi)}{Z}\}^{1/2},
\end{equation}
where $\Xi = 1.2$ takes into account the contribution in
electronic density from 2s- and 3s-  conditions, $\xi = \sigma
Z^{1/3}$ , magnitude $\sigma$ changes from 1 up to 1,15 at
increase Z from 2 up to 80. Various types of I-oscillations in a
crystal connected to various parts of atoms are possible. At
increasing Z the electronic density at the center of atom envelop
is considerably increased when emerge s-electrons of L -, M -, N -
orbitals. Therefore it is expedient to define frequencies of
I-oscillations for atom parts distinguished by the contents
s-electrons. So, the joint oscillations of a nucleus and L -
electrons (nucleus and two 1s electrons) concerning other part of
environment represent $\beta$-type I-oscillations. Elementary
quantum of $\beta$-type I-oscillations may be calculated for atoms
with $Z>2$ with help of the Eq.~(\ref{E19}) at $s
=\sigma(Z-2)^{1/3}$   and $\Xi = 0.2$. It is possible to calculate
elementary quantum of $\gamma$ - type I-oscillations representing
joint oscillation of a nucleus, K - and L - orbitals concerning
other part of an environment for atoms with Z~$>$~8 with help of
the Eq. (\ref{E19}) at $s~=~\sigma(Z-8)^{1/3}$ and $\Xi = 0,056$.
Thus it is possible to describe terms of I-oscillations $\alpha-$,
$\beta-$ and $\gamma-$types by the formula for harmonic oscillator
\begin{equation}\label{E20}
 \hbar\omega_{\nu} = \hbar\omega_{z}(\nu +1/2),
\end{equation}
where oscillatory quantum number $\nu = 0, 1, 2,...$ and $\hbar
\omega_{z}$ is defined by the Eq.~(\ref{E19}). These oscillatory
energy levels together with the origin of energy scale E = 0 are
energy levels of adiabatic crystal model and correspond to
adiabatic $(\alpha -, \beta -, \gamma -$ type) nuclei oscillations
concerning electron system in a crystal. The transitions between
the energy I-oscillatory levels or on the energy I-oscillatory
levels occurring with electron and phonon participation are
non-adiabatic processes which may be studied with the help of
offered adiabatic crystal model. The calculated quantum meanings
depending on Z are shown in Fig. 2 by open points for
$\alpha$-type, $\beta$-type, $\gamma$-type of I-oscillations.
Experimental quantum meanings of I-oscillations for atoms: carbon,
oxygen, sulfur, aluminium, iron, gadolinium, which were measured
by various methods. They are shown in Fig. 2 by closed points.
\begin{figure}
\vspace*{1cm}
\includegraphics[width=8cm]{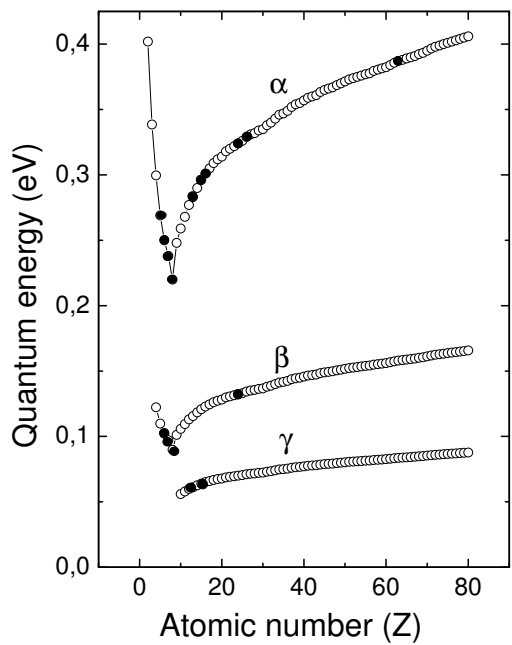}
\vspace*{0cm} \caption {Calculated dependencies of the elementary
quantum energies for I-oscillations of $\alpha$-type ($\alpha$),
of $\beta$-type ($\beta$), and of $\gamma$-type ($\gamma$) from
atomic number Z are shown by open points. Experimental quantum
energies of I-oscillations for several atoms are shown by closed
points.} \label{fig2}
\end{figure}

\section{Creation and destruction of inherent oscillations}

Inherent oscillations can be created at the expense of suitable
energy source for example thermal, optical, energy recombination
in semiconductors. The thermal way is represented improbable, as
the Debye temperatures for inherent $\alpha$-type oscillations
(more than 2500~K) exceed  temperatures of melting for many
materials. The way for excitation of I-oscillations  at the
expense of energy allocated at recombination  in semiconductors
probably is quite effective and practically important when
recombination occurs on the electron-vibrational centers (EVC).
The EVC in semiconductors usually represent an association of
impurity atom with vacancy and are characterized by strong
interaction of electrons with phonons. In the recombination act on
such centers participate  average $S
> 1$ phonons, where S - constant for electron-phonon interaction
(Pekar-Huang-Ryse constant) which can reach several tens. The EVC
equilibrium positions may be changed at recombination. That
promotes to excitation of I-oscillations. Thus, the oscillations
terms of  EVC are shown as deep energy levels. The I-oscillation
of atomic nucleus can formally be characterized by temperature
$T_{\nu} = \hbar\omega_{\nu}/k$ , where k - Boltzmann constant.
$T_{\nu}$ significantly exceed a temperature of a crystal lattice
which are formed by electronic system. Or else, $T_{\nu}$ exceed
temperature of electronic system. EVC disseminates this
oscillatory energy in external environment with the greatest
probability as crystal phonons, I-oscillations and waves of
I-oscillations of the basic substance of a crystal. In result the
recombination on EVC appears as the reason of I-oscillations and
waves connected with phonons and capable  influence on physical
properties, of a crystal. Therefore it was possible to expect
influence of EVC I-oscillations, and also I-oscillations of the
basic substance atoms, on physical properties of semiconductor
crystals and crystal structures.

\section{Description of experiments}

In experiments were used flat polished semiconductor samples by
thicknes 200 mkm containing EVC. It was GaP samples with
impurities of aluminum or sulphur $(\simeq 5\cdot10^{15}
cm^{-3})$: GaP(Al) and GaP(S). Such impurities were chousen
becouse atoms Al and S have masses appreciably exeed mass of atom
Ga. This advantages emergence EVC and generation of
I-oscillations. In experiments also was used quartz samples which
where cut out from a single crystal lengthways normal to axis c,
and also samples of pyrolytic graphite and carbon nanotube films
on substrates which were fabricated by dispersion of graphite by
electronic beam in vacuum \cite{Kos92}. Carbon nanotube films
represent regular structures from carbon nanotube by diameter
about ten nm and in length about 0.1 mkm guided lengthways normal
to a surface of a substrate. Silicon samples with impurities of
phosphorus $(\simeq 5\cdot 10^{15} cm^{-3})$ and oxygen
$(\simeq10^{18} cm^{-3})$: Si(P, O) were used also. In Si samples
EVC was formed by oxygen atoms (A-centers: associations of oxygen
atoms with vacancy, S = 5 for A-centers ). For performance of
electrical measurements  Au and Al contacts to samples were used
with identical success. The contacts rendered by thermal
dispersion of metals in vacuum. Temperature dependencies of
electrical resistivity $\rho(T)$ of samples containing EVC  were
measured within the range from 77~K to 700~K by a method
Van-der-Pau and by two sonde method. The experimental temperature
dependencies of resistivity  in various sites are described by the
formula $\rho(T)=exp(E_{a}/2kT)$. With the help of this formula
were determined the activation energies ($E_{a}$) on various sites
of experimental dependencies $\rho(T)$.

Infra-red reflection (IR) spectra connected with EVC were measured
in the optical range from 15 mkm (83 meV) to 2 mkm (620 meV) at
300K. The angle between IR beam and flat surface of sample was
equal to $45^{0}$. IR reflection spectra was analyzed on the basis
of theory for reflection of classical charged harmonic oscillator
\cite{Hua51, Born54}. IR reflectivity of the oscillator
\begin{equation}\label{21}
  R_{\omega} = \frac{(n-1)^{2} + k^{2}}{(n+1)^{2} + k^{2}}
\end{equation}
is defined by a parameter of refraction (n) and absorption
coefficient (k). They depend on oscillator frequency $(\Omega)$,
frequency of crystal lattice  oscillations $(\omega_{L})$,
coefficient  of damping $(\theta)$, optical permittivity
$(\varepsilon_{opt})$ and optical frequency $(\omega)$ :
\begin{equation}\label{22}
n^{2}-k^{2}=\varepsilon_{opt}+\omega^{2}\frac{\Omega^{2}-\omega^{2}}
{(\Omega^{2}-\omega^{2})^{2}-\omega^{2}\theta^{2}},~~~~~
2nk=\omega_{L}^{2}\frac{\omega
\theta}{(\Omega^{2}-\omega^{2})^{2}+\omega^{2}\theta^{2}}.
\end{equation}

The greatest meanings of  $R_{\omega}$ are located between
$\Omega$ and $\omega_{L}$. Contour of the spectrum depends from
$\varepsilon_{opt}$ and $\theta$. According to the quantum theory
\cite{Ros51,Noz58} the experimental spectra of reflection was
decomposed on components every of which fits spectrum for one
oscillator by selecting parameters $\Omega$, $\omega_{L}$,
$e_{opt}$ and $\theta$. Thus the specified parameters were
defined.

\section{Experimental results and discussion}

\subsection{Temperature dependence of  resistivity}

\begin{figure}
\vspace*{-2cm}
\includegraphics[width=10cm]{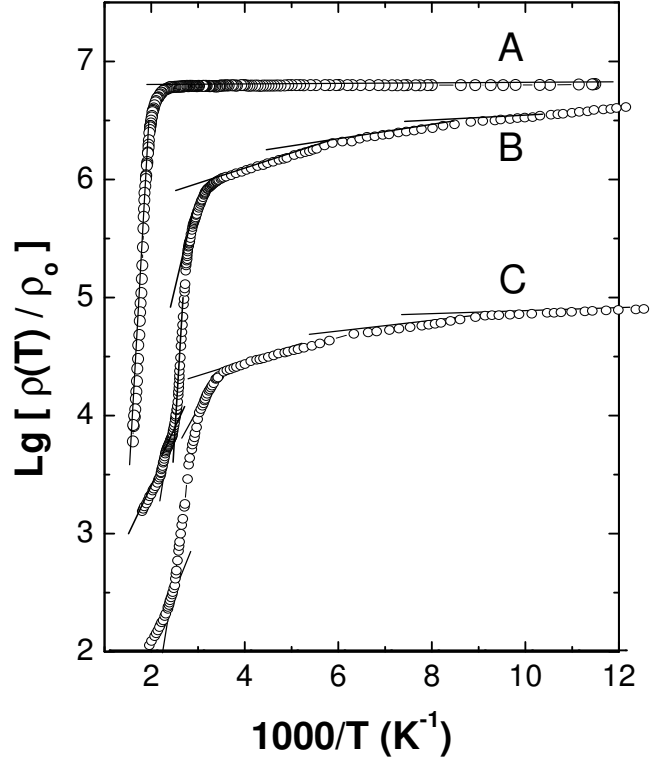}
\vspace*{-1cm} \caption {Temperature dependencies of resistivity
($\rho$) for undoped single crystal GaP (A), for GaP doped by
aluminium (B) and for GaP doped by sulphur (C) are shown by open
circles. Straight lines show tangents to the linear curves sites.}
\label{fig3}
\end{figure}
Typical temperature dependencies of electrical resistivity
$\rho(T)$ for GaP samples are shown in Fig.~\ref{fig3}. Magnitudes
$\rho_{0}$ were selected for each curve that the curves were
conveniently arranged in the Fig.~\ref{fig3}. The sites of curves
with constant inclinations are pointed out by pieces of line
tangent to the curves. The inclination of lines tangents to the
curves relatively axes of coordinates defines magnitudes $E_{a}$.
Magnitudes $E_{a}$, measured at $T \leq 330~K$, are brought in
Table I and basically correspond to known energies of phonons in
GaP \cite{Mars64}.

\begin{table*}[t]
\begin{center}
\caption{Activation energies for GaP(Al) and GaP(S) at $T<330~K$.}
\vspace{0.5cm}
\begin{tabular}{|c|c|c|c|c|c|c|c|c|c|c|}
\hline \hline
\multicolumn{5}{|c|}{Activation energies $E_a$ (meV)
for} & Phonons$^{a}$ &
\multicolumn{5}{|c|}{Activation energies $E_a$ (meV) for }\\
\multicolumn{5}{|c|}{GaP(Al) samples with numbers:} & in GaP &
\multicolumn{5}{|c|}{GaP(S) samples with numbers:}\\
\cline{1-5} \cline{7-11}
\ ~~~~1~~~~ & ~~~~2~~~~ & ~~~~3~~~~ & ~~~~4~~~~ & ~~~~5~~~~ & (meV) & ~~~~6~~~~ & ~~~~7~~~~ & ~~~~8~~~~ & ~~~~9~~~~ & ~~~~10~~~~\\
\hline
\ ~~~8.1 & - & - & 7.2 & 8.0 & - & - & - & - & - &  - \\
\hline
\ ~~15.0 & - & - & 14.8 & 14.5 & 14.25 & 15.0 & 14.8 & - & - & 14.3\\
\hline
\ ~~24.5 & - & - & - & 25.0 & 24.42 & 24.0 & - & 24.6 & - & -\\
\hline
\ - & 35.0 & - & - & - & - & - & - & - & 28.0 & 28.0\\
\hline
\ ~~45.0 & - & - & - & 44.6 & 44.75 & - & 42.0 & - & - & -\\
\hline
\ ~~-~~~ & - & 49.0 & 48.0 & - & 47.00 & 47.0 & - & - & - & -\\
\hline
\ ~~~75.0 & - & 83.0 & - & - & - & - & - & 70.0 & - & -\\
\hline \hline
\end{tabular}
\end{center}
\begin{tabbing}~~~~$^{a}$Reference \cite{Mars64}
\end{tabbing}
\end{table*}

Energy of phonons are introduced in the central column in Table~I.
Other energies in Table~I supposedly  may be connected with
I-oscillations ($\beta$- or $\gamma$-types) of Al or S and
possible with combinations of such oscillations with phonons.

\begin{table*}[t]
\begin{center}
\caption{Activation energies (eV) for GaP(Al) and GaP(S) at
$T>330~K$.} \vspace{0.5cm}
\begin{tabular}{|c|c|c|c|c|c|c|}
\hline \hline
\multicolumn{5}{|c|}{$E_a$  for GaP(Al) samples} & Calculated on & Multiple \\
\multicolumn{5}{|c|}{ with numbers:} &  Eqs.~(\ref{E20}) & $\hbar\omega_{13}$\\
\hline
\ 1 & 2 & 3 & 4 & 5 & ~ & ~\\
\hline
\ ~~0.14~~ & ~~0.14~~ & ~~0.14~~ & ~~0.138~~ & ~~0.137~~ & ~~0.142~~($\nu=0$)& ~\\
\hline
\ 0.28 & 0.29 & 0.29 & 0.28~ & 0.28 & - &~~$\hbar\omega_{13}$~~\\
\hline
\ 0.42 & 0.42 & 0.43 & 0.42~ & 0.43 & 0.425 ($\nu=1)$ & ~\\
\hline
\ - & 0.57 & - & 0.56~ & 0.58 & ~ & 2$\hbar\omega_{13}$\\
\hline
\ 0.71 & - & 0.72 & - & - & 0.707 ($\nu=2$) & ~\\
\hline
\ - & - & - & 0.85 & - & ~ & 3$\hbar\omega_{13}$\\
\hline
\ - & 0.97 & - & - & - & 0.991 ($\nu=3$) & ~\\
\hline
\ 1.10 & - & - & - & 1.11 & ~ & 4$\hbar\omega_{13}$\\
\hline
\multicolumn{5}{|c|}{$E_a$ for GaP(S) samples} & ~ & Multiple \\
\multicolumn{5}{|c|}{ with numbers:} & ~ &$\hbar\omega_{16}$\\
\hline
\ 6 & 7 & 8 & 9 & 10 & ~ & ~\\
\hline
\ 0.15 & 0.15 & 0.15 & ~0.15 & 0.15 & 0.151 ($\nu=0$) & ~\\
\hline
\ 0.30 & 0.29 & 0.30 & 0.30 & 0.31 & ~ & $\hbar\omega_{16}$~\\
\hline
\ 0.60 & - & 0.60 & ~0.61 & - & ~ & 2$\hbar\omega_{16}$\\
\hline
\ - & - & - & - & 0.74 & 0.753 ($\nu=2$) & ~\\
\hline
\ - & - & 0.92 & - & - & ~ & 3$\hbar\omega_{16}$~\\
\hline
\ 1.03 & - & - & - & - & 1.050 ($\nu=3$) & ~\\
\hline \hline
\end{tabular}
\end{center}
\end{table*}

The magnitudes $E_{a}$ for GaP (Al) and GaP (S) measured at $T >
330~K$ are included in Table~II.  The Table~II also contain
energies calculated with help of Eq.~(\ref{E20}) for inherent
oscillations of Aluminum and Sulfur atoms with different $\nu$.
The energies was calculated with $\hbar\omega_{13}$ = 0.283 eV for
Aluminum and $\hbar\omega_{16}$ = 0.301 eV for Sulfur which were
determined in section~III and pointed out in Fig. \ref{fig2}.

One can see from Table~II that activation energies of samples with
each type of impurities can be divided into two groups which
corresponds to two  columns dextral. One group consists of
activation energies which are described by the Eq.~(\ref{E20}).
These activation energies are equal to energy of impurity atoms
I-oscillations and correspond to transitions from oscillatory
energy levels with~$\nu=0,~1,~ 2,~\cdots$ in the minimum of
oscillatory potential where oscillatory energy E = 0. Such
transitions for free quantum harmonic oscillator are forbidden but
they are possible due to interaction of I-oscillator with crystal.
Consequently, inherent oscillators of impurity atoms show duality
of properties that can be explained by their interaction with the
environment. Other group of energies in Table~II consists of ones
multiple $\hbar\omega_{13}$ for Al and $\hbar\omega_{16}$ for S.
These groups of energies are also connected with inherent
oscillations of impurity atoms and correspond to transitions
between oscillator energy levels with different $\nu$. One can see
that $\hbar\omega_{13}$ = 0.283~eV is the same in both groups of
energies for Al impurity, and $\hbar\omega_{16}$ = 0.301~eV is the
same in both groups of energies for S impurity. Consequently, both
groups of energies refer to the same type of centers that show
classical and quantum properties (duality of properties).

The temperature dependencies $\rho(T)$ for pyrolytic graphite (PG)
samples also are characterized by several activation energies. The
magnitudes $E_{a}$ for samples PG are submitted in Table~III. They
were measured at directions of a current parallel to nuclear
planes (L) and perpendicular to nuclear planes (T) of graphite.
Also in Table~III the energies of I-oscillations for carbon are
brought. They calculated with help of the Eq.~(\ref{E20}) for
various $\nu$. These energies was calculated with
$\hbar\omega_{6}$ = 0.25~eV for carbon which was determined in
section~III and pointed out in Fig.~\ref{fig2}.

\begin{table*}[t]
\begin{center}
\caption{Activation energies (eV) of resistivity for graphite
samples at current directions parallel (L) or perpendicular (T) to
atomic planes.} \vspace{0.5cm}
\begin{tabular}{|c|c|c|c|c|c|c|c|c|c|c|}
\hline \hline \multicolumn{9}{|c|}{Activation energies at L and T
current directions}& Calculated & Multiple $$\\
\multicolumn{9}{|c|}{and supposed type of oscillations (Type)}&
on Eqs. (\ref{E20}) & $\hbar\omega_{6}$\\
\hline
\ L & T & L & T & L & T & T & L & Type & - & -\\
\hline
\ ~~-~~ & ~~0.004~ & ~~~~-~~~ & ~~~~-~~~ & ~~~~-~~~ & ~~~~-~~~ & ~0.003~ & ~0.004~ & ~~TA~~ & ~~~~~~~~~~-~~~~~~~~~~ & ~~~~~~~~~-~~~~~~~~~\\
\hline
\ - & - & - & - & - & 0.006 & 0.006 & - & TO & - & -\\
\hline
\ - & 0.029 & 0.021 & - & - & - & 0.024 & - & - & - & -\\
\hline
\ 0.040 & 0.040 & - & - & - & - & - & - & TA & - & -\\
\hline
\ - & 0.044 & - & - & 0.043 & - & - & 0.043 & TA & - & -\\
\hline
\ - & - & - & - & - & - & 0.055 & - & TO & - & -\\
\hline
\ 0.082 & - & 0.082 & 0.084 & - & 0.090 & - & - & LA & - & -\\
\hline \ 0.125 & - & 0.125 & 0.125 & 0.125 & 0.125 & 0.125 &
0.125 & $I_{\alpha}$ & 0.125 ($\nu = 0$) &  -\\
\hline \ 0.250 & 0.250 & 0.247 & 0.245& - & 0.250 & 0.252 & 0.250
&
$I_{\alpha}$ & - & $\hbar\omega_{6}$\\
\hline \ - & - & - & - & - & 0.374 & - & - &
$I_{\alpha}$ & 0.375 $(\nu) = 1$ & -\\
\hline \ 0.500 & - & 0.507 & - & 0.500 & - & - & - &
$I_{\alpha}$ & - & 2$\hbar\omega_{6}$ \\
\hline \ - & - & - & - & - & - & 0.621 & 0.628 &
$I_{\alpha}$ & 0.625 ($\nu$ = 2)& -\\
\hline \ - & - & - & 0.759 & - & - & - & - &
$I_{\alpha}$ & - & 3$\hbar\omega_{6}$ \\
\hline \hline
\end{tabular}
\end{center}
\end{table*}

Top lines of Table~III contain of energies which may be identified
with  phonons of graphite in the certain points of Brillouin zone
appropriate to greater density of phonons. It is possible to
explain energies 0.021 eV, 0.024 eV and 0.029 eV in Table~III by a
combination several acoustical phonons.

Activation energies located in the bottom of Table~III can be
divided into two groups. The first group consist of energies which
close to energies calculated with help of  Eq.~(\ref{E20}). Second
group consist of energies multiple $E_{0}$. The activation
energies may be explained as expense of energy in the acts of
charge carriers generation at transition EVC from a condition with
$\nu$ = 0, 1 and 2 in minimum of potential V($R_{j}$) where
oscillatory energy E~=~0. Others energies are multiple $E_{0}$ and
may be explained by transitions EVC between oscillatory condition
with various $\nu$. These results specify duality properties of
carbon I-oscillator in graphite similar duality  of properties for
I-oscillators in GaP(Al) and GaP(S).

Experimental  activation energies  for  carbon nanotube films on
substrates are given in Table~IV together with the calculated with
help of  the Eq.~(\ref{E20}) data with subject to  I-oscillator
quantum of carbon atom ($\omega_{6} = 0.25$ eV).

\begin{table*}[t]
\begin{center}
\caption{Activation energies (eV) for carbon nanotube films on
quartz  and fluorite  substrates.}
 \vspace{0.5cm}
\begin{tabular}{|c|c|c|c|c|c|c|c|c|c|}
\hline \hline \multicolumn{5}{|c|}{Nanotube films on quartz} &
\multicolumn{3}{|c|}{Nanotube films on fluorite} & Calculated
on & Multiple \\
\multicolumn{5}{|c|}{ substrates with numbers:} &
\multicolumn{3}{|c|}{ substrates with numbers:} & Eq. (\ref{E20}) &  $\hbar\omega_{6}$ \\
\hline
\  1  &  2  &  3  &  4  &  5  &  6  &  7  &  8  & - & - \\
\hline  \ ~0.013~ & ~0.012~ & ~0.013~ & ~~~~-~~~~ & ~~~~~-~~~~~ & ~~~~~-~~~~~ & ~~~~~-~~~~~ & ~~~~-~~~~ & ~~~~~~~~~~~-~~~~~~~~~~~ & ~~~~~~~-~~~~~~~ \\
\hline
\ 0.017 & 0.017 & 0.017 & 0.017 & - & - & - & 0.016 & - & - \\
\hline
\ 0.030 & 0.032 & 0.030 & - & - & 0.029 & - & 0.035 & - & - \\
\hline
\ 0.040 & 0.050 & 0.040 & 0.037 & 0.043 & 0.044 & - & 0.056 & - & - \\
\hline
\ 0.095 & - & - & 0.067 & 0.095 & 0.086 & - & - & - & - \\
\hline
\ 0.127 & 0.125 & 0.125 & - & 0.128 & 0.121 & 0.125 & 0.125 & 0.125 ($\nu$ = 0) & - \\
\hline
\ 0.027 & 0.250 & 0.250 & 0.250 & 0.248 & 0.250 & 0.250 & 0.250 & - &$\hbar\omega_{6}$ \\
\hline
\ - & 0.375 & - & - & - & - & 0.375 & - & 0.375 ($\nu = 1$) & - \\
\hline
\ - & - & - & 0.500 & 0.495 & - & 0.500 & 0.500 & - & $2\hbar\omega_{6}$ \\
\hline
\ - & - & - & - & - & - & ~~0.875~~ & - & 0.875 ($\nu$ = 3) & - \\
\hline \hline
\end{tabular}
\end{center}
\end{table*}

In the top lines of Table~IV are located the energies measured at
$T < 300~K$ which can be identified with phonons and their
combinations. The  bottom lines of Table~IV contain the activation
energies which were measured at $T > 300~K$. This energies can be
divided into two groups. One group consist of the energies which
close to calculated energies for carbon I-oscillations with help
of Eq.~(\ref{E20}). It gives the certain basis to connect them to
transitions of carbon I-oscillators from condition with $\nu = 0,
1, 2, 3$ in a condition with oscillatory energy E = 0. Other group
consist of energies multiple $\hbar\omega_{6} = 0.25$ eV which can
be connected with transitions of the same I-oscillators between
oscillatory condition differing by $\nu$ on 1 or 2.

The results of measurements and analysis of activation energies in
samples Si(P, O) are qualitatively identical to the stated
results, but the  activation energies at $T > 300~K$ close to
energies of oxygen I-oscillations  (in   A-centers) at $\nu = 0,
1, 2, 3, 4$.

It is impossible to explain the experimental dependencies
$\rho(T)$ with scattering of charge carriers by phonons at $T <
300~K$ since the  scattering is capable to create an opposite
effect to the observed effect of reducing resistance when
increasing the temperature. We connect experimental activation
energies with creation of free charge carriers at the expense of
oscillations energy of EVC for example EVC created by atoms Al or
S in GaP, by atoms C in graphite  or in carbon nanotube films, and
by atoms O (A-centers) in silicon. The coincidence of calculated
and experimental activation energies  at $T > 300~K~$ allows us to
connect them with I-oscillatory terms of impurity atoms forming
EVC. These terms are shown as deep energy levels.

\subsection{Infrared spectra}

\begin{figure}
\vspace*{0cm}
\includegraphics[width=8cm]{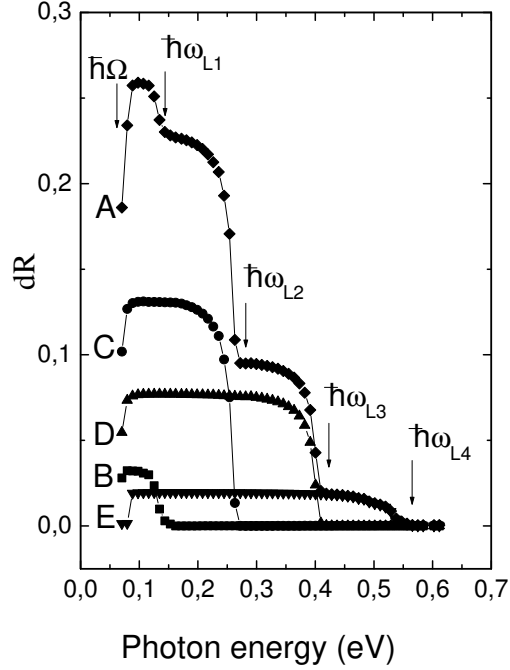}
\vspace*{-0.5cm} \caption {Spectrum of reflectivity change (dR)
which caused by impurity Al in GaP (A) and components of the
spectrum connected with different  I-oscillations of Al (B, C, D,
E). Oscillator energy ($\hbar\Omega$) and energies of Al
$\alpha$-type I-oscillations in different oscillations states
$\hbar\omega_{Li}~(i = 1, 2, 3, 4)$ are shown by arrows.}
\label{fig4}
\end{figure}

 Typical spectrum of reflectivity change
(dR) in GaP(Al) is shown in Fig.~\ref{fig4} by curve A.
The dR caused by introduction of impurity atoms (Al) in GaP  in
concentration $\simeq 5 \cdot 10^{15} cm^{-3}$. According to the
theory \cite{Ros51, Noz58} the given spectrum was decomposed into
components, which in Fig. 4 are denoted as B, C, D, E. The sum of
the calculated spectra B, C, D and E, coincides with the
experimental spectrum A if energies $\hbar\omega_{Li}$ (i = 1, 2,
3, 4 ) coincide with energies of Al $\alpha$ -type I-oscillators:
$0.5 \hbar\omega_{13}$, $\hbar\omega_{13}$, $1.5
\hbar\omega_{13}$, $2 \hbar\omega_{13}$; $\hbar\omega_{13}$ =
0.283~eV. It is possible to see from these data that Al
I-oscillators show duality of properties because in optical
transitions the energy level of "zero oscillations" $(0.5
\hbar\omega_{13})$ is shown. It is possible to connect energy
$\hbar\Omega$ = 61 meV with $\gamma$- type of atom Al
I-oscillations. The attenuation $(\theta)$ in I-oscillators is
great: $(\theta/\Omega)$ = 0.09. The best consent between
experimental and calculated spectra is reached if
$\varepsilon_{opt} = 2$ though for  GaP $\varepsilon_{opt} =
8.457$ \cite{Klei60}. It may be the $\varepsilon_{opt} = 2$ is
necessary to carry not to a crystal GaP, but only to EVC. Thus the
section of photon capture by EVC can depend on wavelength of
phonons cooperating with EVC. The similar results were received at
research of EVC reflectivity spectra  formed by impurity atoms of
sulfur for which $\hbar\omega_{16} = 0.301~eV$.

\begin{figure}
\vspace*{0cm}
\includegraphics[width=8cm]{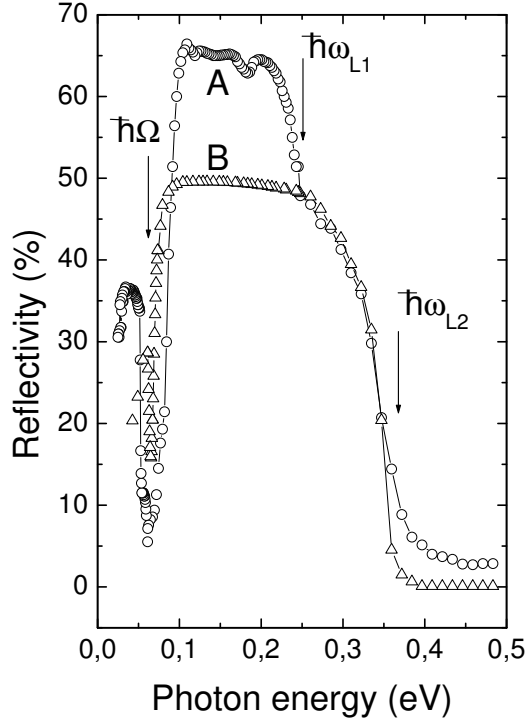}
\vspace*{-0.5cm} \caption {Reflectivity spectrum of carbon
nanotube film on molybdenum substrate (A) and  calculated (at
$\hbar\omega_{L2}$ = 0.375 eV, $\nu$ = 1)  reflectivity spectrum
for carbon $\alpha$-type I-oscillator (B). Oscillator energy
($\hbar\Omega$) and energies of carbon $\alpha$-type
I-oscillations are shown by arrows.} \label{fig5}
\end{figure}

IR reflectivity spectrum  of carbon nanotube film on molybdenum
substrate (curve A) is shown in Fig.~\ref{fig5}.
Spectrum A was decomposed on  reflection bands of  two oscillators
with energies: $\hbar \Omega = 60$ meV,  $\hbar \omega_{L1} = 0.25
eV$ and $\hbar\omega_{L2} = 0.375 eV$. The first of the energies
is equal to elementary quantum of carbon $\beta$-type
I-oscillations, second energy is equal to quantum of carbon atom
$\alpha$-type I-oscillations ($\hbar\omega_{6}$)  and
 last of the energies is equal to carbon $\alpha$-type
I-oscillations at $\nu = 1$. Thus, the experimental data about
interaction $\alpha$- and $\beta$-types of I-oscillations  and
their influence on  IR reflection spectra were received.

The typical reflectivity spectrum  of carbon nanotube film on a
copper substrate contains one band which is well described by the
calculated reflectivity spectrum of  oscillator with parameters
$\hbar\Omega = 60$ meV and $\hbar\omega_{L} = 0,25$ eV which are
accordingly equal to elementary quantums  $\beta$- and
$\alpha$-type I-oscillations of carbon atom.

\begin{figure}
\vspace*{-1cm}
\includegraphics[width=8cm]{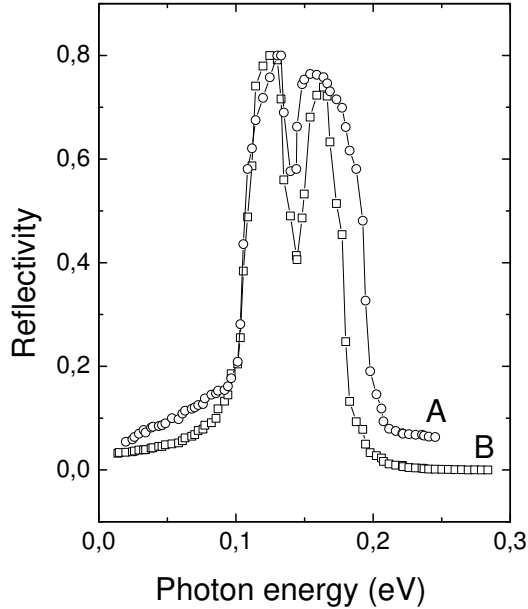}
\vspace*{-0.5cm} \caption {Reflectivity spectrum of   quartz (A)
and calculated  reflectivity spectrum of oxygen $\alpha$-type
I-oscillator (B).} \label{fig6}
\end{figure}

Reflectivity spectrum of  single-crystal quartz  is shown in
Fig.~6 (curve A). Curve B in Fig.~\ref{fig6} represents calculated
reflectivity spectrum of oxygen $\alpha$-type I-oscillator with
the following parameters: $\hbar\Omega = 0.11$ eV;
($\omega_{L}/\Omega) = 0.25$ and $\varepsilon_{opt} = 1.2$,
$(\theta/\Omega) = 0.011$. One can see that energy 0.11 eV is
equal to $\hbar\omega_{8}$/2 at $\nu$ = 0 in Eq.~(\ref{E20}).
Frequency $\omega_{L}$ can be identified with frequency of a
phonon. The satisfactory consent of spectra A and B in a Fig.~6
confirms the electron-vibrational nature of the given reflection
band and allows definitely to connect her with excitation of
$\alpha$-type I-oscillations of oxygen atom at $\nu$ = 0. Some
distinction of spectra A and B is explained by the contribution to
reflection a components with $\omega_{L} = 2\Omega$ and
$\omega_{L} = 3\Omega$.

\begin{figure}
\vspace*{0cm}
\includegraphics[width=12cm]{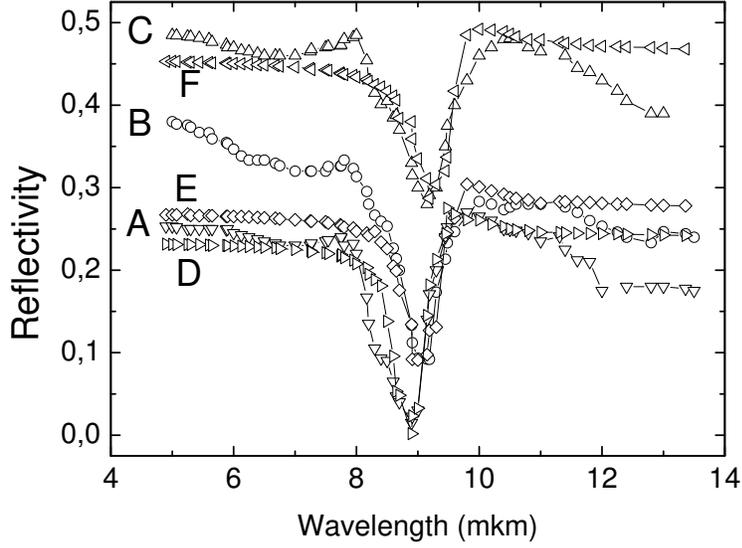}
\vspace*{0cm} \caption {Experimental reflectivity spectra of
carbon nanotube film on quartz substrate after film fabrication
(A), two month later fabrication (B), six months later fabrication
(C) and reflectivity spectra of oxygen $\alpha$-type I-oscillator
(D, E, F), calculated at meanings of $\varepsilon_{opt}$ equal to
2, 20, 250, accordingly.} \label{fig7}
\end{figure}
Typical experimental reflectivity spectra of  carbon nanotube film
on quartz substrate are submitted in Fig.~\ref{fig7} (curves A, B,
C). Spectrum A is measured at once after cultivation of a film.
Spectra B and C were measured 2 and 6 months later accordingly.
Between measurements the film was stored at room conditions. The
IR reflection factor ($R_{\omega}$) increase considerably during a
storage of samples. The similar changes of reflection spectrum for
carbon nanotube film occur at her heating in vacuum. The changes
of a spectrum occurring during two months of storage in room
conditions can be achieved at heating of a samples in vacuum at T
$\simeq$ 500~K approximately within one hour. It allows to connect
changes of the spectrum with changes of film structure. The
experimental spectra contain a minimum in that area where the
maximum of quartz reflection band is located. This reflection
maximum is characteristic for impurity oxygen in silicon  where it
is displaced from 0.13 eV to 0.11 eV at increase of oxygen
concentration  and formation of quartz disseminations. This
reflection maximum  is characteristic also for silicon oxides, and
for quartz. The reflection maximum   definitely can be connected
to $\alpha$-type I-oscillations of oxygen atom at $\nu$ = 0.
Occurrence of reflection minimum close to 0.11 eV caused by carbon
nanotube film on quartz substrate cannot be explained by
anti-reflection coating action of a film because of the reflection
minimum  is observed at various thickness of a film and substrate.
It is possible to explain this minimum by interaction between
I-oscillators  of oxygen atoms in quartz substrate and carbon
atoms in nanotube film. Curves D, E, F in Fig.~\ref{fig7}
represent spectra of oscillator reflectivity calculated at
parameters $\hbar\Omega$ = 0.11~eV and $\hbar\omega_{L}$ = 1.25~eV
with $\varepsilon_{opt}$ equal to 2, 20 and 250 accordingly. The
first from specified energies coincides with energy of
$\alpha$-type I-oscillations for oxygen atom at $\nu$ = 0 in
Eq.~(\ref{E20}) ($\hbar\omega_{8}/2$ = 0.11~eV) but the second
energy is multiple to elementary quantum of $\alpha$-type
I-oscillations for carbon atom ($\hbar\omega_{6}$ = 0,25~eV). The
reflectivity spectra of nanotube films on quartz after thermal
treatment  are satisfactorily described by oscillator spectrum  at
$\varepsilon_{opt} = 10^{3} - 10^{4}$ but characteristic energy
$\hbar\Omega$ = 0.11~eV and $\hbar\omega_{L}= 1.25$~eV remain
constant. It allows to explain the given experimental spectra by
interaction with each other of oxygen and carbon I-oscillations at
participation of phonons.

\begin{figure}
\vspace*{-1cm}
\includegraphics[width=9cm]{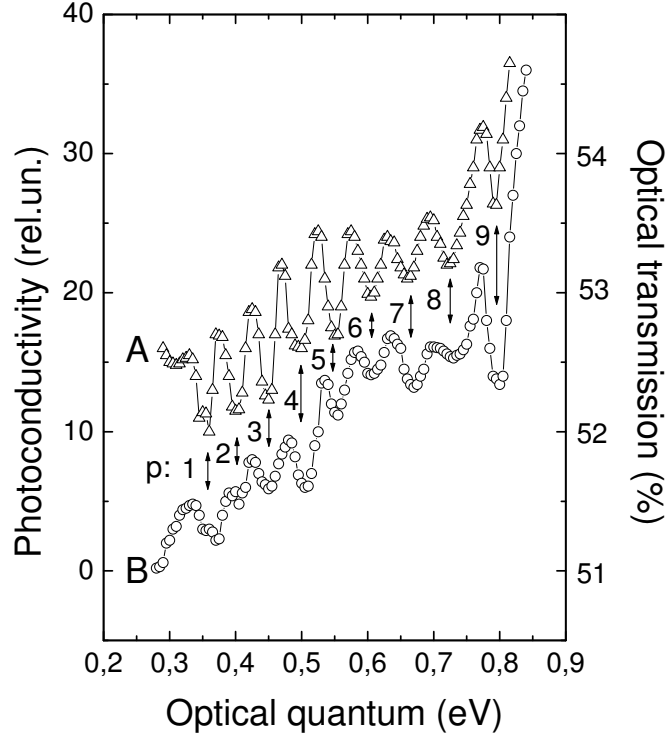}
\vspace*{-1cm} \caption {Photoconductivity spectrum (A) and
spectrum of IR transmission (B) measured at 80~K and connected
with electron transitions on A-centers in silicon sample. Number
of phonons which participate in electron-vibrational transitions
is denoted as p.} \label{fig8}
\end{figure}

In silicon samples the IR reflectivity spectra, optical
transmission and photoconductivity at electron-vibrational
transitions on A - centers have phonon structure and are described
by line Pekar-Huang-Rhyse spectra \cite{Pec52, Hua50}. Typical
photoconductivity  (curve A) and experimental spectra of IR
transmission (curve B) for containing A-centers ($10^{14}
cm^{-3}$) Si single-crystal are introduced in Fig.~\ref{fig8}.
Every of spectra lines correspond to participation of p phonons.
The energies of electron transitions without phonons participation
(p = 0) determined on basis of the theory \cite{Pec52, Hua50}
coincide with energies of $\alpha$- type oxygen atom
I-oscillations counted from top of valence energy zone in silicon.
These spectra are measured in polarized IR radiation when the IR
electrical vector was directed as normal to [100] in a crystal.
The spectra contain extremes which differ from each other on
energies multiple energy of optical phonon in Si ($\simeq 55$
meV). Height  of spectral lines follows dependence ($S^{p}/p!$),
constant of electron-phonon interaction (Pekar-Huang-Rhyse
constant) S~$\approx$~5. Energy  of electron transitions without
phonons participation (p = 0) is close to energy $E_{v}+ 0.33$ eV
(where $E_{v}$ - top of a valence energy zone) which correspond to
$\nu = 1$ for oxygen $\alpha$- type I-oscillator. At others
polarizations of IR radiation it is possible to allocate spectra
with participation others (optical and acoustic) phonons with the
same meanings of constant S and energies of electron transition
with p = 0 conterminous with I-oscillatory terms of oxygen atom.
Curve B in Fig.~\ref{fig8} represents typical for EVC spectrum of
negative photoconductivity. Curve B in Fig.~\ref{fig8} represents
spectrum of IR transmition. One can see that the increase of IR
transmission (that is reduction of absorption) corresponds to
photoconductivity increase. Thus terms of oxygen I-oscillations in
A-center is similar to other atoms I-oscillations in structure
EVC. The I-oscillation terms are shown as deep energy levels  in
semiconductors.

\subsection{Discussion}

Inherent  oscillations of atomic nuclei relatively electron system
(relatively electron envelops of atoms) in crystals and molecules
exist  that one can see  from results of experimental and
theoretical researches of different authors. The equations system
Eqs.~(\ref{E14})-(\ref{E17}) at n~=~0 describes oscillations of
one crystal cell  that is oscillations of appropriate molecule. In
correspondence with Eq.~(\ref{E18}) the oscillatory spectrum of a
two-nuclear molecule contains frequencies of nuclei I-oscillations
when electron envelops are motionless (on I-oscillations
frequencies) and nuclei move relatively  each other with inverse
phase. Such oscillations really are available in spectra of
two-nuclear molecules. F. Vilesov and M. Akopiyan with help of
optical mass spectrometry method investigated the ionization
spectra for hundreds of molecules in various oscillatory
conditions \cite{Vil69}. In particular they  defined the energies
of transitions from the adjacent oscillatory conditions in
molecule of oxygen ($O_{2}$) whose differences are close to
quantum of oxygen $\alpha$-type  I-oscillations (0.22~eV). That
coordinates with I-oscillations presence in two-nuclear molecules.

G. Pastore and E. Smargiassi  have applied in Ref. \cite{Pas91}
Car-Parrinello \cite{Car85} and Born-Oppenheimer \cite{Ha92}
molecular dynamics for calculation model system consisted of eight
silicon atoms forming the periodic diamond lattice. Car-Parrinello
dynamic basically takes into account conditions with various
meanings of oscillatory quantum number $\nu$, but Born-Oppenheimer
dynamic corresponds to the minimal oscillatory energy of a nucleus
($\nu = 0$). The comparison of application results for both
dynamics allows to define transition energies between conditions
with various meanings $\nu$. In the Car-Parrinello dynamic
"fictitious masses" about some hundreds of electron mass are used.
They are inertia parameters assigned to every orbital degrees of
electron freedom at a nucleus motion. It is obvious that
fictitious mass of electron envelop in every atom sufficiently
exceed mass of nucleus. Therefore at movement of a nucleus the
electronic envelop remain motionless and changeless. It
corresponds to inverted adiabatic approximation (inverted in
relation to well-known adiabatic approximations) when nuclei move
(hot nuclei) but the electrons "do not heat up" systematically in
the presence of the hot nuclei (cold electrons). It corresponds to
the representation about massive and motionless electronic envelop
(on frequencies of nucleus I-oscillations) which we used for
definition energy quantums of I-oscillations.

In agreement with Ref. \cite{Pas91} the density of electronic
conditions in considered atomic model system reaches the maximum
near 1.3 eV that close to width of the forbidden zone in Si single
crystal ($E_{g} = 1.16~eV$ at temperature 0~K). The minimal energy
in the calculated electron spectrum  (0.6~eV) exceeds the maximal
phonons energy ($\cong 55\cdot10^{-3}$~eV) that satisfies
inequalities Eqs.~(\ref{E10})-(\ref{E11}) and does improbable
systematical flow of energy between nuclei and electrons. The
applicability  of adiabatic approximation is thus reasonable. In
\cite{Pas91} also are calculated: kinetic energy of the electrons
($T_{e}$), electronic energy ($V_{e}$), kinetic energy of nuclei
($T_{z}$), conserved energy ($E_{cons} = T_{e} + T_{z} + V_{e}$),
physical total energy ($E_{phys} = E_{cons} - T_{e}$). Energies
$E_{cons}$ and $E_{phys}$ are strictly constant. Energies $V_{e}$
and $T_{e}$ change on 0.43~eV and $0.81\cdot10^{-3}$~eV
accordingly in opposite phase to each other with the common period
$\tau_{1} = 2.15\cdot10^{-14}$ s which correspond to energy $E_{1}
= ( 2\pi\hbar/\tau_{1}) = 0.194$~eV. The specified energy 0.43 eV
can be connected to transitions of silicon I-oscillator between
condition in minimum of potential $V(R_{j})$ which coincide with
oscillatory energy  of nucleus E = 0 and condition with $\nu = 1$
that is calculated with help of Eq. (\ref{E20}) for silicon atom
with meaning $\hbar\omega_{14} = 0.29$~eV. If it is correct then
calculations in \cite{Pas91} proves observable on experience the
duality properties of I-oscillators. In accordance with
\cite{Pas91} the electron system acts on a nucleus by force which
in Car-Parrinello approach $(F_{CP})$ and in Born-Oppenheimer
approach $(F_{BO})$ oscillate in  phase with the common period
$\tau_{2} = 4.32\cdot10^{-13} s$ which correspond to energy $E_{2}
= (2\pi\hbar /\tau_{2}) = 9.67\cdot10^{-3}~eV$. Besides, the
difference $(F_{CP} - F_{BO})$ oscillate with two periods which
correspond to energies $\cong~9.672\cdot10^{-3}~eV$ and $\cong~
0.2998~eV$. Last of these energies is near to energy  of
I-oscillator transitions for silicon atom between the adjacent
oscillatory conditions with $\nu = 0$ (Born-Oppenheimer approach)
and $\nu = 1$ (Car-Parrinello approach).

One can see from these results that in adiabatic approximation the
complete crystal energy  is constant but there is the oscillating
process of energy exchange  between  system of electrons and
system of nuclei. Therefore  the adiabatic conditions
Egs.~(\ref{E10})-(\ref{E11}) concerning systems of electrons and
nuclei is correct only for average energy on an time interval
exceeding the periods $\tau_{1}$ and $\tau_{2}$. The
I-oscillations of atomic nuclei carry out important role in
oscillating process of energy exchange between systems of
electrons and nuclei in crystals. These I-oscillations give raise
to such phenomena as phonon drag of electrons \cite{Arx99},
thermal superconductivity, and hypercoductivity \cite{Arx00} at
very high temperatures.

\section{Conclusion}

Electronic envelop of any atom represents the new collective
quality of electron system  in comparison with quality  of free
electrons. The electronic envelops of different atoms can unite
with each other and form crystal lattices  due to that the nuclei
of atoms get an opportunity to carry out adiabatic inherent
oscillations (I-oscillations) relatively electronic system of a
crystal. In adiabatic approach there is no systematic energy flow
from electronic system  to nuclei system or in opposite direction.
However there is the oscillating process of energy  exchange
between system of nuclei and system of electrons which is
submitted by inherent oscillations in particular. These
I-oscillations are a nucleus oscillations relatively motionless
(on I-oscillation frequencies) nucleus environment in a crystal
($\alpha$-type of I-oscillations). It  can be also oscillations of
a nucleus together with K-electrons ($\beta$-type of
I-oscillations) or nucleus together with K- and L-electrons
($\gamma$-type of I-oscillations) relatively  motionless  (on
frequency of I-oscillations) environment. Each atom in adiabatic
model of crystal  is submitted as appropriate I-oscillator.
Non-adiabatic processes can be considered as transitions between
stationary terms of I-oscillators.

Stationary I-oscillations and also waves of I-oscillations can
exist in crystals. The energy spectrum of I-oscillations and waves
of I-oscillations can be described by the formula for harmonic
oscillator (at neglecting by non-harmonicity). The elementary
quantum of harmonic I-oscillation depend on nuclear number, and
for $\alpha$-type I-oscillations they considerably exceed energy
of phonons.

 The effective creation of I-oscillations and waves of such oscillations
can be carried out at the expense of recombination energy of
electrons and holes on electron-vibrational centers in crystals.
I-oscillations energy terms of the electron-vibrational centers
are shown  as deep energy levels in semiconductors.

The I-oscillations influence on physical properties of crystals
and crystal structures. Thus the inherent oscillations
representing oscillations of atomic nuclei relatively electronic
system  in crystals or molecules are the important physical
reality.

\begin{acknowledgments}
I thank Dr. Z. Ya. Kosakovskaya for carbon nanotube films on
substrates given for researches.
\end{acknowledgments}

\end{document}